\documentclass{moriond}

\usepackage{amsfonts}

\def\be{\begin{equation}}
\def\ee{\end{equation}}
\def\bea{\begin{eqnarray}}
\def\eea{\end{eqnarray}}

\newcommand{\Photo}{}

\begin{document}
\vspace*{4cm}
\title{MINKOWSKI-TENSOR-BASED SHAPE ANALYSIS METHODS ON THE SPHERE}

\author{ C. COLLISCHON${}^1$, M. KLATT${}^2$, C. RÄTH${}^3$, M. SASAKI${}^1$ }

\address{${}^1$Dr. Karl Remeis-Sternwarte, FAU Erlangen-Nürnberg; ${}^2$Institut für Theoretische Physik II, HHU Düsseldorf; ${}^3$Deutsches Zentrum für Luft- und Raumfahrt (DLR)}

\maketitle\abstracts{
Recently, Minkowski Tensors (MT) have gained popularity for morphological analysis tasks. As opposed to the scalar Minkowski functionals (MF; in 2D given by area, perimeter and Euler characteristic), MT can characterize symmetry and orientation of a body.
This has been used for a variety of tasks, e.g. to detect interstellar bubbles by tracing back the origins of filaments in HII-regions, or to search for alignment of structures in the CMB.
I present a marching-square-based method for calculating MT and MF on the sphere for maps in the Healpix format. MT are calculated for a local neighborhood and can then be summed up/averaged over a larger region, using their additivity property. 
This provides the possibility of localized analyses looking for CMB anisotropies and non-Gaussianities at varying scales. }

\section{Background}
Minkowski functionals (MF) and tensors (MT) are powerful and versatile shape descriptors. In 2D, the MF are up to prefactors given by area, perimeter, and Euler characteristic. Their tensorial counterparts can be defined the following way, using the position $\vec r$ and the normal vector $\vec n$:

        Let $K$ be a convex shape, then
            \begin{eqnarray}
            W_0^{a,0}(K) :=&&\int_K \vec{r}^a\, \mathrm dr\\
            W_\nu^{a,b}(K) :=&&\int_{\partial K} \vec{r}^a\otimes \vec{n}^b\, \lambda_\nu\, \mathrm dr
            \end{eqnarray}
            with $\nu\in\{1,\, 2\}$, $\lambda_1=1$, $\lambda_2=\kappa$ and $a,b \in \mathbb{N} _0$ and 
            \be
            (\vec{r}^a\otimes\vec{n}^b)_{i_1\ldots i_{a+b}} =\frac{1}{(a+b)!}\sum_{\sigma\in S_{a+b}} r_{i_{\sigma(1)}}\ldots r_{i_{\sigma(a)}}\cdot n_{i_{\sigma(a+1)}} \ldots n_{i_{\sigma(a+b)}}\, ,
            \ee
            where $S_n$ is the permutation group of $n$ elements

		They are additive $\left( W_i^{j,k}(K) + W_i^{j,k}(L) = W_i^{j,k}(K\cup L) -W_i^{j,k}(K\cap L) \right)$ for convex bodies $K, L$\,, which allows generalizations to certain non-convex shapes, including shapes defined by pixel images. For more properties, see e.g. Schröder-Turk et al. (2011)\cite{schroederturk2011}.
		
		Hadwiger's theorem states that any additive, continuous, and translation invariant functional on convex bodies can be expressed as a linear combination of MF \cite{hadwiger1957}. A similar theorem by Alesker states that the MT contain all additive morphological properties of a shape \cite{alesker1999}.
		These definitions hold for the Euclidean plane; extension to the sphere is subject of current research.
		
		\section{Minkowski maps}
		
\begin{figure}
		\centering
		\includegraphics[width=0.27\textwidth]{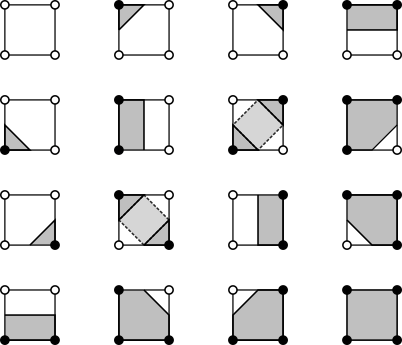}
		\hspace{3ex}
		\includegraphics[width=0.6\textwidth]{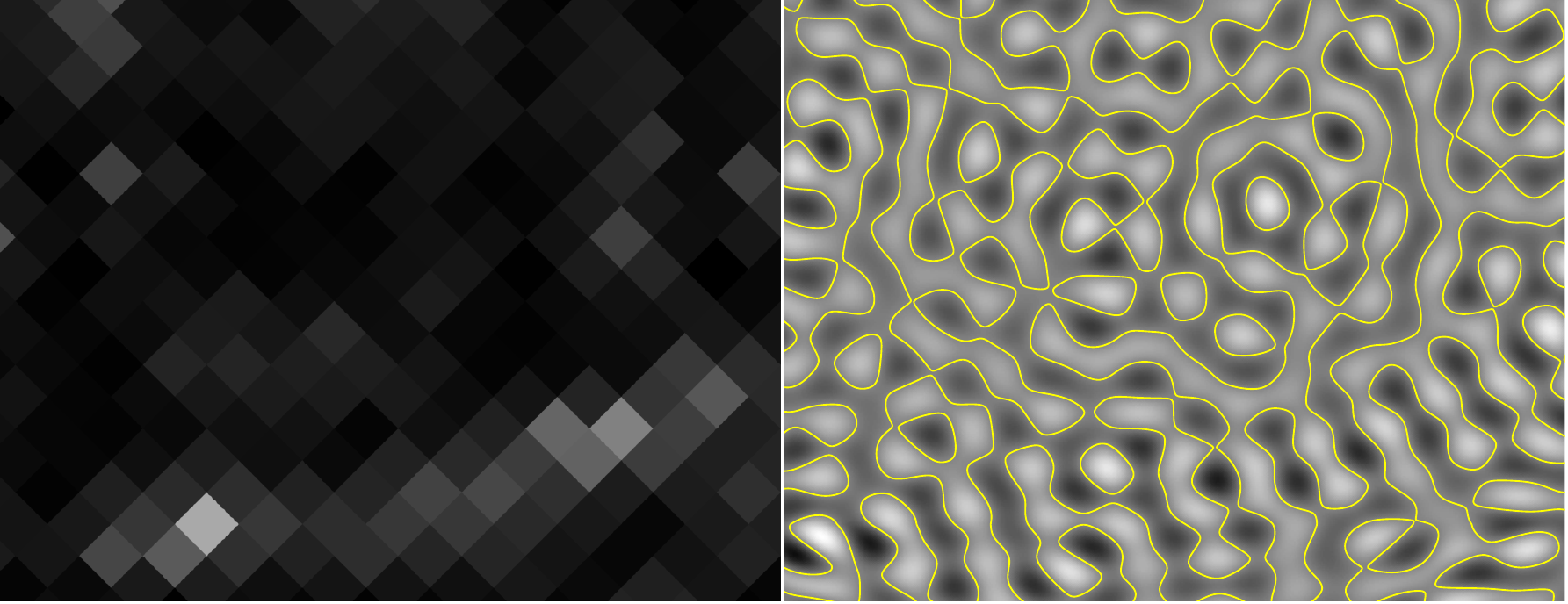}
		\caption{Left: The 16 possible configurations of a marching square window. In the diagonal cases a choice must be made whether the shape should be connected. The exact corner position is interpolated. Center/right: Elongation as measured by the ratio of eigenvalues of $W_1^{0,2}$ (left) and the underlying Gaussian field with contours used for the Minkowski map (right).}
		\label{marching}
		\end{figure}

		Minkowski tensors can be calculated for pixel images with a range of grayscale values using a marching square technique. A 2x2\,px-region is considered, where pixels above a chosen threshold are seen as part of the body. The 16 possible configurations of a 2x2\,px-window are shown in Fig.~\ref{marching}. The desired MF/MT is calculated for the resulting simple shape in this window. To obtain the MF/MT of a larger region, the MF/MT of all contained marching squares are summed up, using the additivity property. 
		The result of this can be displayed as a Minkowski map with each map pixel describing the shape of a local input image region, as shown in Fig.~\ref{marching} (right).
		
		\section{Application}
		
		Raw MT need to be turned into scalars for visualization and analysis. This is possible in many ways using, e.g., eigenvalues, traces, or directions. 
      
      	Shape properties can, e.g., be measured by the ratio of eigenvalues of $W_1^{0,2}$. This is shown in Fig.~\ref{marching} (right), where parts of a Gaussian field that happen to have elongated structure can be seen as having a large ratio of eigenvalues.
		As opposed to previous MT/MF-analyses of the cosmic microwave background, which considered the whole sky at once (e.g. Joby et al. 2019 \cite{joby2019}), Minkowski maps enable localized search for anisotropies and non-Gaussianities. 
		More generally, Minkowski maps have been successfully used to automatically detect bubble-like structures in the Magellanic Clouds \cite{collischon2021}.


\section*{References}

\bibliography{references}

\end{document}